\documentclass[twocolumn,showpacs,preprintnumbers,amsmath,amssymb,prb]{revtex4-1}
\usepackage{graphicx}% Include figure files
\usepackage{dcolumn}% Align table columns on decimal point
\usepackage{bm}% bold math
\usepackage{color}

\begin{document}
%\nofiles
%\preprint{APS/123-QED}
%\title{\textcolor{red}{Enhancement and suppression of macroscopic
%quantum tunneling due to the Josephson Leggett mode}}
%\title{Theory of macroscopic quantum tunneling with Josephson-Leggett
%collective excitations in multi-band superconducting Josephson junctions}
%\title{Variation of macroscopic quantum tunneling rates via
%Josephson-Leggett modes}
%\title{Josephson-Leggett-mode-assisted macroscopic quantum tunneling in
%Josephson junctions with two-band superconductors}
%\title{Theory of macroscopic quantum tunneling with inter-band
%fluctuations in multi-band superconducting Josephson junctions}
\title{Theory of macroscopic quantum tunneling with Josephson-Leggett collective excitations in multi-band superconducting Josephson junctions}

\author{Hidehiro Asai$^{1}$} 
\email{hd-asai@aist.go.jp}
\author{Yukihiro Ota$^{2}$}
\email{otayuki@riken.jp} 
\author{Shiro Kawabata$^{1}$}
\author{Masahiko Machida$^{2}$} 
\author{Franco Nori$^{3,4}$} 
\affiliation{$^{1}$Electronics and Photonics Research Institute
(ESPRIT),  National Institute of Advanced Industrial Science and
Technology (AIST), Tsukuba, Ibaraki 305-8568, Japan}
\affiliation{$^{2}$Center for Computational Science and e-Systems (CCSE), Japan Atomic Energy Agency, Kashiwa, Chiba 277-8587, Japan}
\affiliation{$^{3}$Center for Emergent Matter Science (CEMS), RIKEN, Wako-shi, Saitama 351-0198, Japan}
\affiliation{$^{4}$Physics Department, University of Michigan, Ann Arbor, Michigan 48109-1040, USA}

\date{\today}

\begin{abstract}
Collective excitations reveal fundamental properties and potential
 applications of superconducting states.  
We theoretically study macroscopic quantum tunneling (MQT) in a
 Josephson junction composed of multi-band superconductors, focusing on
 a phase mode induced by inter-band fluctuations: the Josephson-Leggett (JL)
 collective excitation mode. 
Using the imaginary-time path-integral method, we derive a formula for
 the MQT escape rate for low-temperature switching events.
We clarify that the JL mode has two major effects on the MQT: (i) the
 zero-point fluctuations enhance the escape rate, and (ii) the quantum
 dissipation induced by the couplings to the gauge-invariant phase
 difference suppresses the quantum tunneling. 
We show that the enhancement exceeds the suppression for a wide range of junction parameters. 
This enhancement originates from the single-mode interaction between
 the tunneling variable and the inter-band fluctuations. 
\end{abstract}

\maketitle

\section{Introduction}
Josephson junctions show phenomena caused by macroscopic-scale
quantum coherence and non-linear dynamical properties. 
Their unique properties come from couplings between
superconducting gauge-invariant phase differences and the
electromagnetic field, leading to practical applications, as
seen, e.g., in Ref.~\onlinecite{Barone}. 
Macroscopic quantum tunneling
(MQT)~\cite{CLMQT,Simanek,generalMQT1,generalMQT2} is one of the
characteristic phenomena of Josephson junctions. 
Superconducting-to-resistive switching events in current-biased
Josephson junctions are related to MQT at low
temperatures, where thermal excitations are negligible. 
A wide variety of junction systems (artificial niobium-based
junctions~\cite{1stMQT,EMMQT}, grain boundary junctions~\cite{Bauch;Claeson:2005,Bauch;Lombardi:2006}, and intrinsic Josephson
junction stacks in single-crystalline high-$T_{\rm c}$ cuprate  
superconductors~\cite{Inomata;Kawabata:2005,Jin;Meuller:2006,Kubo;Takano:2012}) 
show this tunneling phenomenon. 
The theoretical aspects have been studied well, depending on the types of
Josephson junctions.~\cite{Kato;Imada:1996,Kawabata;Tanaka:2005,Machida;Koyama:2007,Savelev;Nori:2007,Sbochakov;Nori:2007,NoriLongMQT,Savelev;Nori:2008,KawabataMQT,2gapOtaMQT}
MQT in Josephson junctions plays an important role in
Josephson phase qubits and the relevant superconducting quantum
engineering.~\cite{pqubit1,pqubit2,pqubit3,pqubit4,You;Nori:2011} 
Hence, studying MQT in Josephson junctions attracts a great deal of
attention theoretically and experimentally, to find quantum
characteristics of superconducting devices. 

The discovery of superconducting materials, including magnesium
diboride~\cite{1stMgb2} and iron-based compounds,~\cite{1stIron} triggered
the studies on multi-band superconductivity. 
These superconductors have intriguing properties, originating from the
multiple superconducting gaps opening in different parts of the Fermi
surfaces.~\cite{IronReview,IronReview2,Mgb2Cp,Mgb2Raman,Mgb2Review}
Notable Josephson effects are prediced in junctions with multi-band
superconductors.~\cite{2gapAgter,Onari;Tanaka:2009,Linder;Sudbo:2009,Linder;Sperstad;Sudbo:2009,Inotani;Ohashi:2009,2gapOtaPRL,Golubov;Dolgov:2009,2gapOtaMQT}
The characteristic behaviors in these systems originate from the presence
of multiple gauge-invariant phase differences coupled by 
inter-band Josephson
coupling.~\cite{Leggett:1966,Tanaka:2001,2gapOtaPRL,Sharapov;Beck:2002,Gurevich;Vinokur:2003,Iskin;SadeMelo:2006}  
Specifically, a phase mode induced by inter-band fluctuations,
which is referred to as a Josephson-Leggett (JL)
mode,~\cite{2gapOtaPRL} can lead to singular behaviors. 
However, the JL-mode excitations are not coupled directly to the
electric field, owing to their neutral-superfluid
feature~\cite{Sharapov;Beck:2002,Iskin;SadeMelo:2006}.  
Instead, they interact with the Josephson plasma (JP) mode (i.e., in-phase
motion of
superfluids).~\cite{2gapOtaPRL} 
Thus, a careful and systematic study of the interaction between the
JP and the JL modes is desirable for exploring
characteristic phenomena in Josephson junctions composed of
multi-band superconductors. 

In this paper, we construct a theory of MQT in a Josephson junction
formed by a conventional single-band superconductor and a two-band
superconductor (a hetero Josephson junction), to clarify the
effects of the JL mode on low-temperature switching events in Josephson
junctions. 
From theoretical considerations of the dynamics of gauge-invariant phase
differences, we choose a tunneling path along the center-of-mass motion
of the phase differences, and consider the inter-band fluctuations to be
the environment for the center-of-mass motion. 
We evaluate the MQT escape rates, varying different junction parameters. 
We show that the inter-band fluctuations have both positive and negative
effects on the MQT. 
Zero-point fluctuations of the JL mode enhance the MQT escape rate,
whereas the quantum dissipation induced by the JL mode suppresses the quantum
tunneling. 
The former was found by two of the authors (YO and
MM),~\cite{2gapOtaMQT} focusing only on a specific junction parameter. 
Thus, the present approach successfully extends the previous results in 
Ref.~\onlinecite{2gapOtaMQT}, and reveals two distinct features of the
JL mode, i.e., amplification and reduction. 
Moreover, we show that the zero-point-fluctuation enhancement
exceeds the quantum-dissipation suppression. 
Therefore, we find that the escape rate is significantly enhanced by the
JL mode.  
In these junctions, the dissipation effect is marginal 
because there is only one dissipation channel corresponding to a
monochromatic JL-mode. 
We also examine the dependence of the escape rate on the inter-band
Josephson energy for junction parameters which are typical for  
$\textrm{BaFe}_2 \textrm{As}_2$ and $\textrm{MgB}_2$.
We find that the effects of inter-band fluctuations on MQT strongly
depend on the nature of the JL mode characterized by the superconducting
material parameters of the junction. 

This paper is organized as follows.
In Sec.~II, we introduce a minimal model of multi-band Josephson junctions. 
In Sec.~III, we describe a theory of MQT in this Josephson junction and
derive the MQT escape-rate formula. 
In Sec.~IV, we evaluate the escape rate for various junction
parameters. 
Section V presents a summary. 
\begin{figure}
\includegraphics[width=7.2cm]{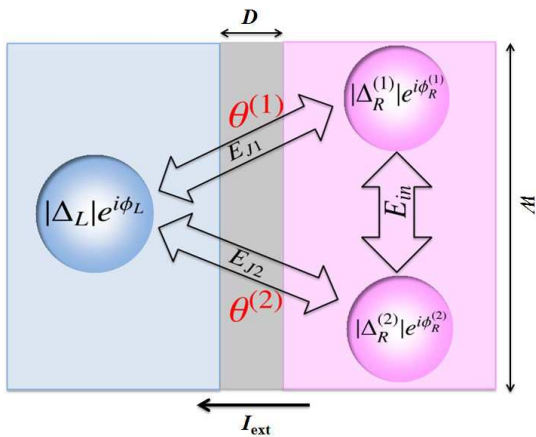}
\caption{(Color online) Schematic diagram of a Josephson junction composed of a
 single-gap superconductor (left electrode) and a two-gap superconductor
 (right electrode). Two gauge-invariant phase differences
 $\theta^{(1)}$ and $\theta^{(2)}$ are defined between the electrodes.} 
\end{figure} 

\section{Model}
We study a minimal model of a multi-band Josephson junction, as ssen in
Fig.~1, to find an esssential feature of the JP-JL coupling. 
The system is composed of a conventional single-gap superconductor and
a two-gap superconductor. 
The superconducting electrodes are separated by an insulating layer
with thickness $D$ and area $W$.  
The DC current $I_{\rm ext}$ is applied to this junction. 
The right electrode has two gaps, 
$|\Delta^{(1)}_{\rm R}| e^{i \phi^{(1)}_{\rm R}}$ 
and $|\Delta^{(2)}_{\rm R}| e^{i \phi^{(2)}_{\rm R}}$, 
whereas the left electrode has a single gap 
$|\Delta_{\rm L}| e^{i \phi_{\rm L}}$. 
These superconducting phases are coupled to each other, via the
Josephson couplings $E_{\rm J1}$, $E_{\rm J2}$, and $E_{\rm in}$. 
The standard Josephson energy associated with Cooper-pair tunneling
between the superconducting electrodes is characterized by $E_{\rm J1}$
and $E_{\rm J2}$. 
The inter-band Josephson energy associated with the tunneling between 
the two bands is characterized by $E_{\rm in}$.~\cite{Leggett:1966}

Now, we show two key variables in this paper, the center-of-mass phase
and the relative phase.  
Using the gauge-invariant phase differences $\theta^{(1)}$ and
$\theta^{(2)}$ between the electrodes (See Fig.~1), the center-of-mass
phase is 
\begin{equation}
\theta 
=
\frac{\alpha_2}{\alpha_1 + \alpha_2} \theta^{(1)} 
+ \frac{\alpha_1}{\alpha_1 + \alpha_2} \theta^{(2)}, 
\end{equation} 
with a dimensionless constant $\alpha_{i}$, related to the density of
states of the $i$th-band electron near the interface between the right
electrode and the insulator.~\cite{2gapOtaMQT}
The relative phase is 
\begin{equation}
 \psi = \theta^{(1)} - \theta^{(2)}. 
\end{equation}
When the voltage difference is $V$, the Josephson
relation is~\cite{2gapOtaPRL} 
\begin{equation}
\frac{\partial \theta}{\partial t}
=
\frac{2eD}{\hbar} V,
\end{equation}
with the electric charge $e$ and the Planck constant $\hbar$. 
The Josephson relation indicates that $\theta$ is directly coupled to 
the electric field, but $\psi$ does not. 
In this paper, we focus on a short Josephson junction, that is, $D$ is
much smaller than the Josephson penetration depth. 
Hence, we ignore the spatial modulation of $\theta^{(1)}$ and
$\theta^{(2)}$, 
and the influence of solitonic excitations shown in
Refs.~\onlinecite{NoriLongMQT,Sbochakov;Nori:2007}. 
%Refs.~\cite{NoriLongMQT,Sbochakov;Nori:2007}.
\section{Formulation}
%\begin{figure*}
%\includegraphics[width=12cm]{Fig2.eps}
%\caption{(Color online) 
%Schematic diagrams of two effects of the Josephson-Leggett mode on a
% macroscopic quantum tunneling (MQT) process. (a) MQT-enhancement by the
% decrease of the barrier height. (b) MQT-suppression by dissipation
 %(quantum dissipation). }
%\end{figure*}
\begin{figure}
\includegraphics[width=7.5cm]{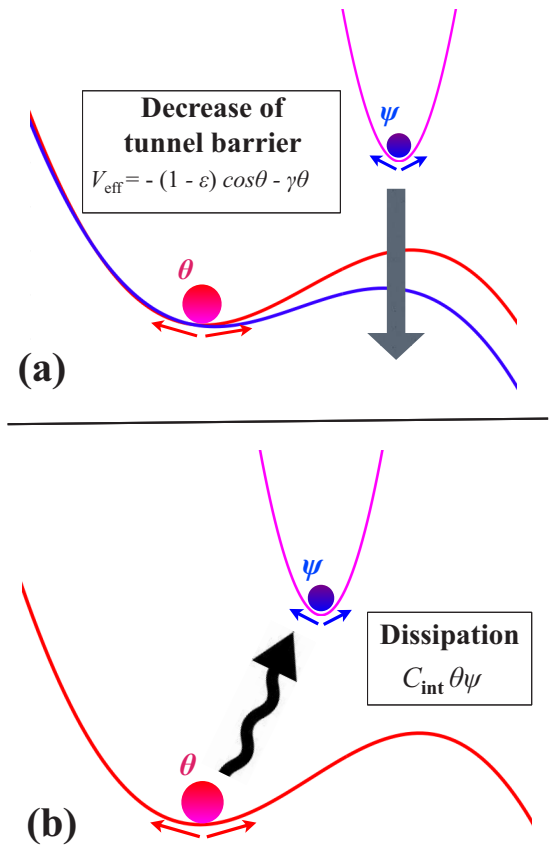}
\caption{(Color online) 
Schematic diagrams of two effects of the Josephson-Leggett (JL) mode on a
 macroscopic quantum tunneling (MQT) process. (a) MQT-enhancement by the decrease of the barrier height. The curves indicate the potential energy with (blue) and without (red) JL mode. 
 (b) MQT-suppression by dissipation (quantum dissipation). The wiggling arrow schematically shows the energy dissipation from $\theta$ to $\psi$ via linear coupling $C_{{\rm int}} \theta \psi$.}
\end{figure}

We formulate the MQT escape rate, based on the semiclassical
approximation with the imaginary-time path-integral
method.~\cite{Simanek}
Our discussion is divided into four steps.  
First, we show the Lagrange formalism of our junction, useful for the
path-integral method. 
Second, from a physical point of view, we find a plausible tunneling
path for low-temperature switching events in the present junction. 
Our approach is to choose a specific path and reduce the issue into an
effective 1D tunneling problem. 
Third, we show that the relevant Euclidean (imaginary-time) Lagrangian
can be mapped into the Caldeira-Leggett model.~\cite{CLMQT}
In this step, we mainly use a small $\psi$-expansion. 
Finally, we obtain the MQT escape rate, using a technique based on the
influence-functional method.~\cite{Simanek,CLMQT,KawabataMQT}

\subsection{Real-time Lagrangian}
The real-time effective Lagrangian $\mathcal{L}$ for our junction
is~\cite{2gapOtaPRL} 
\begin{eqnarray}
\mathcal{L} (\theta, \psi)
&=& 
\frac{1}{2} m_{\rm cm} 
\Bigl( 
\frac{d \theta}{d t} \Bigr)^2
+
\frac{1}{2} m_{\rm rlt} 
\Bigl( 
\frac{d \psi}{d t} \Bigr)^2 
+E_{\rm J1} \cos{\theta^{(1)}} \nonumber \\
&& 
+E_{\rm J2} \cos{\theta^{(2)}}
+E_{\rm in} \cos{\psi}+E_{\rm J} \gamma\, \theta, \label{basicL}
\end{eqnarray}
with 
\mbox{$m_{\rm cm} = \hbar^2/2E_{\rm c}$} and 
\mbox{$m_{\rm rlt} = \hbar^2/2 (\alpha_1 + \alpha_2)E_{\rm c}$}. 
The charging energy is denoted by $E_{\rm c}$. 
The total Josephson energy coupling is 
$E_{\rm J}=E_{\rm J1} + E_{\rm J2}$. 
The dimensionless bias current is $\gamma = I_{\rm ext}/I_{\rm c}$. 
The critical current $I_{\rm c}$ is related to $E_{\rm J}$. 

In this paper, we take a positive value for the inter-band Josephson coupling
 $E_{\rm in}$.~\cite{2gapOtaPRL}
Thus, we focus on the case of a $0$ phase shift in the
two-band superconductor 
(i.e., $\phi_{\rm R}^{(1)}-\phi_{\rm R}^{(2)} \equiv 0\,\mod 2\pi$).  
We can find that our results do not change qualitatively,
when a $\pi$ phase shift (i.e., $\pm s$-wave) occurs. 

\subsection{Determining a tunneling path}
We now seek a predominant tunneling path on $(\theta,\,\psi)$. 
In this paper, we choose an {\it in-phase tunneling path} along the
$\theta$-axis. 
Here, we justify this choice based on the physical properties of
$\theta$ and $\psi$. 
First, $\theta$ has a direct coupling to the electric field,
 whereas $\psi$ does not. 
Thus, the switching event in this junction is caused by 
the tunneling of $\theta$.
This tunneling is strongly enhanced by the bias current $\gamma$ 
because the potential barrier height along the $\theta$ axis decreases
with increase of $\gamma$. 
Second, $\psi$ tends to be fixed to $0$ or $\pi$, because 
the dynamics of the relative phase is subjected to a restoring force 
induced by the inter-band coupling (In this paper, $\psi$ tends to be $0$
because $E_{\rm in} > 0$).   
Although $\psi$ fluctuates around these fixed values, the amplitude of
the fluctuations is relatively small compared to the oscillation of
$\theta$, as described below. 
Moreover, the fluctuations are not affected by the bias current
$\gamma$, in contrast to the tunneling of $\theta$. 
Therefore, the switching event in a high-bias current condition 
$\gamma \simeq 1$ may occur, via the tunneling along the $\theta$-axis. 
Thus, we can reduce our issue to an effective 1D tunneling problem along
this in-phase path.

Let us more closely examine the dynamical behaviors of $\theta$ and
$\psi$ around the in-phase tunneling path. 
The amplitudes of $\theta$-oscillations and $\psi$-fluctuations are 
characterized by, respectively, $m_{\rm cm}^{-1}$ and 
$m_{\rm rlt}^{-1}$. 
Since the dimensionless constants $\alpha_{1}$ and $\alpha_{2}$ are
small ($\alpha_{i} < 1$), we have small fluctuations of $\psi$. 
Hence, we examine the switching event in this Josephson junction, using
the in-phase tunneling path with small relative-phase fluctuations. 
Along this tunneling path, the dynamics of $\theta$ is
expressed by a particle under the so-called washboard potential. 
Futthermore, the dynamics of $\psi$ can be expressed by a simple
harmonic oscillator, with angular frequency 
\begin{equation} 
\omega_{\rm L} = \frac{1}{\hbar} 
\sqrt{2(\alpha_1 + \alpha_2) E_{\rm c} E_{\rm in}},  
\end{equation}
In other words, $\psi$ is regarded as a bosonic {\it environment} for the
center-of-mass phase. 
We will show these points, via the derivation of the Euclidean Lagrangian
with the expansion of $\mathcal{L}$ around the in-phase tunneling path. 

\subsection{Euclidean Lagrangian around an in-phase tunneling path}
Now we derive the Euclidean Lagrangian, with three steps. 
Throughout this paper, we denote the imaginary time as $\tau\,(= it)$. 
First, we take a small $\psi$-expansion, up to second order. 
We obtain the Euclidean Lagrangian
\mbox{
\(
\mathcal{L}^{\rm E} 
=
\mathcal{L}^{\rm E}_{\rm cm} 
+ 
\mathcal{L}^{\rm E}_{\rm rlt} 
+  
\mathcal{L}^{\rm E}_{\rm int}
\)}, with 
\begin{eqnarray}
\mathcal{L}^{\rm E}_{\rm cm} 
&=& 
\frac{m_{\rm cm}}{2}
\Bigl( \frac{d \theta}{ d \tau}\Bigr)^2 
- 
E_{\rm J} \bigl( \cos{\theta} + \gamma \theta  \bigr), \\
\mathcal{L}^{\rm E}_{\rm rlt} 
&=&  
\frac{m_{\rm rlt}}{2}
\Bigl( \frac{d \psi}{ d \tau}\Bigr)^2  
+ 
\frac{1}{2} m_{\rm rlt}\omega_{\rm L}^{2} \psi^2, \\ 
\mathcal{L}^{\rm E}_{\rm int} 
&=& g_{+} E_{\rm J} \psi^2 \cos{\theta} - g_{-} E_{\rm J} \psi \sin{\theta}.
\end{eqnarray}
The first term ($\mathcal{L}^{\rm E}_{\rm cm}$) is the Lagrangian for
the center-of-mass $\theta$, the second term 
($\mathcal{L}^{\rm E}_{\rm rlt}$) is for the relative phase $\psi$, and
the third term describes the interaction between $\theta$ and $\psi$. 
The coupling constants $g_{+}$ and $g_{-}$ are 
\begin{eqnarray} 
g_{+} &=& \frac{1}{2 E_{\rm J}} \Bigl[ \
 \frac{\alpha_1^2}{(\alpha_1+\alpha_2)^2} E_{\rm J1} +
 \frac{\alpha_2^2}{(\alpha_1+\alpha_2)^2} E_{\rm J2}  \Bigr],  \\
g_{-} &=&  \frac{1}{E_{\rm J}} \Bigl[ \ 
 -\frac{\alpha_1}{(\alpha_1+\alpha_2)} E_{\rm J1} + 
 \frac{\alpha_2}{(\alpha_1+\alpha_2)} E_{\rm J2}  \Bigr]. \label{gpm}
\end{eqnarray}
We find that $g_{+}$ is positive, whereas $g_{-}$ vanishes when the
parameters of the respective gaps are equivalent: $\alpha_1 = \alpha_2$ and
$E_{\rm J1} = E_{\rm J2}$. 

Second, in order to remove the non-linearity with respect to $\psi$ in
the interaction Lagrangian, we use the mean-field
approximation.~\cite{2gapOtaMQT}
The expectation values of $\psi$ and $\psi^{2}$ for the ground state
(i.e., zero-temperature limit) of 
$\mathcal{L}^{\rm E}_{\rm rlt}$ are 
\begin{equation}
\langle \psi \rangle_{\psi} =0,
\quad
\langle \psi^{2} \rangle_{\psi} 
= \frac{\hbar}{2m_{\rm rlt}\omega_{\rm L}}
\equiv \Omega^{2},  \label{mfapsi}
\end{equation}
where the symbol $\langle \cdot \rangle_{\psi}$ indicates the
expectation value with respect to $\psi$.
Using these values, we rewrite  $\mathcal{L}^{\rm E}_{\rm int} $ as a
summation of the expectation values and the deviation from them, 
\(
\mathcal{L}^{\rm E}_{\rm int} = 
\langle \mathcal{L}^{\rm E}_{\rm int} \rangle_{\psi}  
+ \delta \mathcal{L}^{\rm E}_{\rm int}
\). 
Omitting higher-order fluctuations, we obtain the linearized interaction
Lagrangian, 
\begin{eqnarray}
\mathcal{L}^{\rm E}_{\rm int} 
= g_{+} \Omega^2 \cos{\theta} - g_{-} \psi \sin{\theta}.
\end{eqnarray}
We then derive the effective Euclidean Lagrangian with the mean-field
approximation, 
\begin{equation}
\mathcal{L}^{\rm E} (\theta, \psi) 
=
 \frac{m_{\rm cm}}{2} \Bigl( \frac{{\rm d} \theta}{{\rm d} \tau}\Bigr)^2 
+ 
V_{\rm eff} (\theta) 
+\mathcal{L}^{\rm E}_{\rm rlt}
- g_{-} \psi \sin{\theta},
\label{eq:Lagrangian_mfa} 
\end{equation}
with 
\begin{equation}
V_{\rm eff} (\theta) 
= - E_{\rm J} \bigl[ (1-\varepsilon) \cos{\theta} 
+ \gamma \theta  \bigr] \label{veff}, 
\end{equation}  
where $\varepsilon = g_{+} \Omega^2$.  

Third, we expand Eq.~(\ref{eq:Lagrangian_mfa}) around the local
minimum of $V_{\rm eff}$, denoted by 
\(
\theta_{0} =\arcsin[{\gamma/(1-\varepsilon)}]
\). 
In this paper, we focus on the case $\gamma \simeq 1$; this is typical
for MQT experiments. 
After performing a constant phase-shift transformation, 
which does not change the path-integral measure, we obtain 
\begin{equation}
\mathcal{L}^{\rm E} (\theta, \psi)
\approx
 \frac{m_{\rm cm}}{2} 
\Bigl( \frac{d \theta}{d \tau}\Bigr)^2 
+ 
\widetilde{V}_{\rm eff} (\theta)
+ 
\mathcal{L}^{\rm E}_{\rm rlt}
- C_{\rm int}\theta \psi + \delta V , \label{mfaL}
\end{equation} 
with
\begin{equation}
\widetilde{V}_{\rm eff}(\theta) = 
\frac{\hbar^{2}\omega_{\rm eff}^{2}(\gamma)}{4E_{\rm c}}
\left(
\theta^{2} - \frac{\theta^{3}}{\theta_{1}}
\right), 
\end{equation}  
where $\theta_{1}=\cot\theta_{0}$ and 
$C_{\rm int} = E_{\rm J}g_{-}\cos\theta_{0}$.
We have dropped constants irrelevant to $\theta$ and $\psi$. 
The current-dependent Josephson-plasma frequency is 
\begin{equation}
\omega_{\rm eff}(\gamma)
= 
\frac{
\sqrt{2E_{\rm c}E_{\rm J}(1-\varepsilon)}
}{\hbar}
\left[
1- \left(\frac{\gamma}{1-\varepsilon}\right)^{2}
\right]^{1/4} .
\end{equation}
We stress that the effect of the JP-JL coupling explicitly appears in
this formula.  
Furthermore, we find that the interaction term 
$C_{\rm int}\theta \psi$ is essentially the same as the system-bath
interaction in the Caldeira-Leggett model~\cite{CLMQT}. 
The last term in eq.~(\ref{mfaL}) is the counterterm~\cite{Simanek} 
\(
\delta V = (C_{\rm int}^{2}/2m_{\rm rlt}\omega_{\rm L}^{2})\theta^{2}
\), which is added for reproducing Hooke's law between $\theta$ and
$\psi$ [i.e., $(\theta-\psi)^{2}$]. 

\subsection{Escape rate formula}
We now show the formula for the MQT escape rate $\Gamma$. 
At the low-temperature limit, the escape rate~\cite{Simanek} is 
\(
\Gamma = (2/\hbar\beta) {\rm Im}\,K(\beta)  
\), 
with the inverse temperature $\beta$ and 
\begin{eqnarray}
K(\beta)
&=&
\int_{-\infty}^{\infty} d\psi 
\int_{\theta(0)=0,\psi(0)=\psi}^{\theta(\beta)=0,\psi(\beta)=\psi}
\mathcal{D}\theta(\tau) \mathcal{D}\psi(\tau) \, \nonumber \\
&&\times \! \exp \left\{
-\frac{1}{\hbar}\int_{0}^{\hbar\beta}
d \tau \,\widetilde{\mathcal{L}}^{\rm E}[\theta(\tau),\psi(\tau)]
\right\}.
\end{eqnarray}
One of the authors (SK)~\cite{KawabataMQT} developed a method to
evaluate the MQT escape rate, for this class of Lagrangian. 
This approach is essentially the same as the influence-functional
method.~\cite{Simanek,CLMQT}
Thus, we find that when $\beta \to \infty$ (i.e., zero-temperature
limit) for the case $\gamma \simeq 1$
\begin{eqnarray}
&&
\Gamma
= 
\omega_{\rm eff} \sqrt{\frac{30 S_{\rm B} }{\pi \hbar}}  
\Bigl(  1 + \frac{S_{\rm D} }{2 S_{\rm B} }    \Bigr) 
\exp{\Bigl[ - \frac{1}{\hbar} (S_{\rm B} +S_{\rm D}) \Bigr]} ,  \label{escape} 
\end{eqnarray}
with 
\begin{eqnarray}
&&
S_{\rm B} 
= 
\frac{8}{15} \frac{\hbar^2}{2E_{\rm c}} \omega_{\rm eff} \, \theta_{1}^2 , 
\quad  \\
&&
S_{\rm D} 
=
\frac{8 \pi C_{\rm int}^{2} \theta_1^2}
{m_{\rm rlt} \omega_{\rm L}^2 \omega_{\rm eff}} 
\int_{0}^{\infty} g(z)\,dz,
\end{eqnarray}
where $g(z)$ means the effects of the memory kernel in terms of the
influence functional method, 
\begin{equation}
g(z) 
=
\frac{ z^4}{  
[ (\omega_{\rm L} / \omega_{\rm eff})^{2} + z^2  ] \sinh^2{(\pi z)}} .  
\end{equation}
Here, $S_{\rm B}$ is the bounce action of the tunneling particle $\theta$
along the extremal path on the potential 
$\widetilde{V}_{\rm eff}(\theta)$, 
whereas $S_{\rm D}$ is the dissipative action, which
corresponds to the energy dissipation from $\theta$ to the environment
$\psi$.  

Before closing this subsection, let us summarize the role of the JL mode on
MQT based on our theory described above.
On the one hand, the zero-point fluctuations give a positive non-zero
$\varepsilon$.  
This quantity effectively reduces the tunneling barrier height, as seen
in Eq.~(\ref{veff}). 
As a result, the zero-point fluctuations enhance the MQT escape rate.
On the other hand, the linear interaction $C_{\rm int} \theta \psi$ in
Eq.~(\ref{mfaL}) causes energy dissipation from $\theta$ to the
environment $\psi$.  
The effect of this {\it quantum dissipation} appears as the dissipative
action $S_D$ in Eq.~(\ref{escape}), then it suppresses the MQT escape
rate. 
In Figs. 2 (a) and (b), we show the schematics of these two major roles
of the JL mode. 

\section{Escape rates with different junction parameters}
\begin{figure}
\includegraphics[width=8cm]{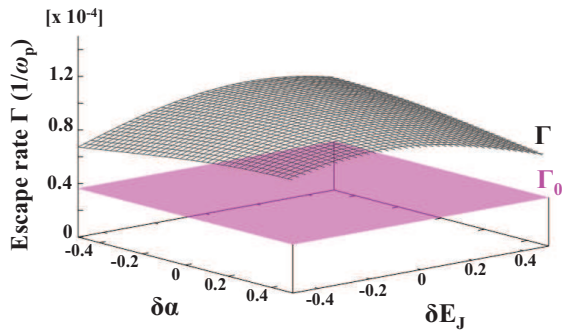}
\caption{(Color online) Macroscopic quantum tunneling escape rate,
 versus the density-of-states difference $\delta \alpha$ and the
 Josephson-energy difference $\delta E_{\rm J}$ between the two
 tunneling channels shown in Fig.1.  $\omega_p  = \sqrt{2 E_{\rm C} E_{\rm J} }/\hbar$ is the Josephson-plasma frequency.
 The black mesh surface ($\Gamma$)
 shows the  escape rate with the inter-band fluctuations. 
In contrast, the purple mesh surface ($\Gamma_{0}$) indicates the escape
 rate without the inter-band fluctuations.
 %The black mesh and purple surface indicates the escape rate with $\Gamma$ and without $\Gamma_0$
 %the inter-band fluctuations 
  }
\end{figure}
\begin{figure}
\includegraphics[width=7.5cm]{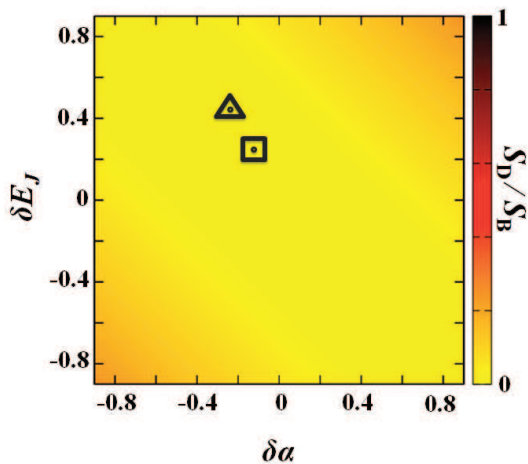}
\caption{(Color online) Ratio of the dissipative action to the bounce
 action, $S_{\rm D}/S_{\rm B}$. 
 The open square and triangle correspond to the parameters for the iron-based
 superconductor  $\textrm{BaFe}_2 \textrm{As}_2$ and $\textrm{MgB}_2$,
 respectively.} 
\end{figure}
\begin{figure}
\includegraphics[width=7.5cm]{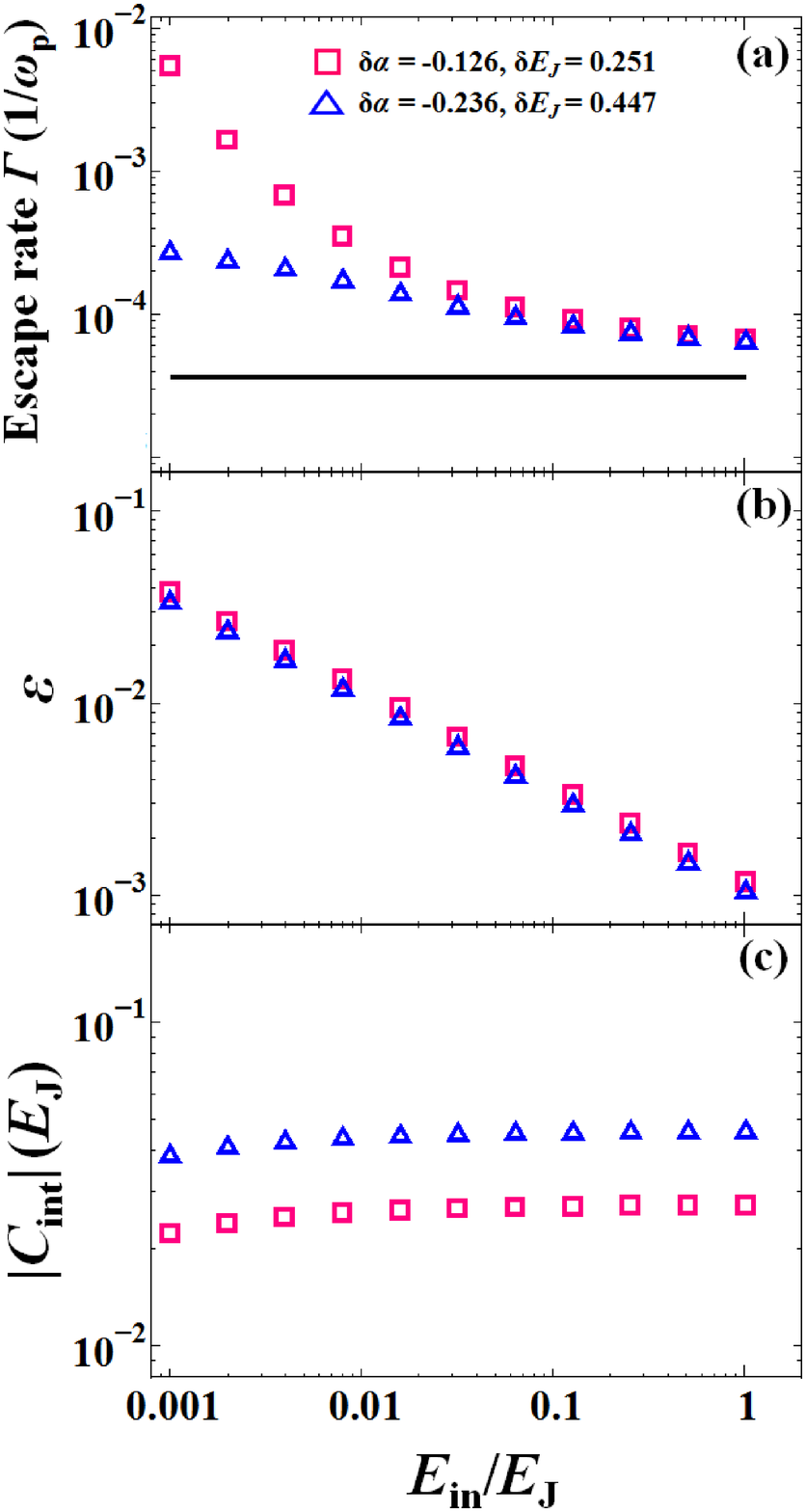}
\caption{(Color online) (a) Macroscopic quantum tunneling escape rate, 
(b) parameter $\varepsilon = g_{+} \Omega^2$ for the magnitude of the zero-point fluctuations of the Josephson-Leggett mode (see Eq.~(14)), 
 and (c) parameter $C_{\rm int} = E_{\rm J}g_{-}\cos\theta_{0}$ for the coupling strength between the
 tunneling particle $\theta$ and the environment $\psi$ (see Eq.~(15)), 
 with different inter-band Josephson couplings,
 $E_{\rm in}/E_{\rm J}$. 
 The $\varepsilon$ and $|C_{\rm int}|$ contribute to the enhancement and the suppression of the quantum tunneling, respectively.
 The horizontal solid line in (a) indicates the escape rate without 
 inter-band fluctuations. The open square and the triangle indicate the
 results for  
$(\delta \alpha = -0.126,\, \delta E_{\rm J} = 0.251)$ and 
$(\delta \alpha = -0.236,\, \delta E_{\rm J} = 0.447)$, respectively. }
\end{figure}
We now numerically evaluate the MQT escape rate (\ref{escape}). 
In order to discuss the MQT in general 2-band Josephson junctions, we
perform the calculations for different junction parameter sets 
\mbox{$(\alpha_{i},\,E_{{\rm J}i})$}, with fixed  
\(
E_{\rm c}/E_{\rm J}\,(=0.002)
\), 
\(
\gamma\,(=0.9)
\), 
and 
\(
\alpha_{1}+\alpha_{2}\, (=0.1)
\). 
We use two parameters characterizing the differences between the two
tunneling channels, 
\begin{equation}
\delta \alpha 
=  \frac{\alpha_{1} - \alpha_{2}}{\alpha_{1}+\alpha_{2}}, \quad
\delta E_{\rm J}
= \frac{E_{\rm J1} - E_{\rm J2}}{E_{\rm J}}. 
\end{equation} 
The former characterizes the density-of-states difference in the vicinity
of the interface, while the latter is the normalized Josephson-energy
difference. 
We examine $\Gamma$ in the parameter space 
\mbox{$(\delta \alpha,\, \delta E_{\rm J})$}. 

Figure 3 shows the MQT escape rate as a function of $\delta \alpha$ and
$\delta E_{J}$, for $E_{\rm in}/E_{\rm J} = 0.1$. $\omega_p  = \sqrt{2 E_{\rm C} E_{\rm J} }/\hbar$ is the Josephson-plasma frequency.
The black mesh indicates the escape rate with the JL mode,
while the purple surface indicates the {\it bare} escape rate
$\Gamma_0$, namely $\Gamma$ without the JL mode 
(i.e. $C_{\rm int}=\varepsilon=0$). 
At the origin of the \mbox{$(\delta \alpha,\, \delta E_{\rm J})$} space in
which all band parameters are equivalent, $C_{\rm int}$ becomes zero. 
Therefore, the MQT suppression by quantum dissipation does not appear at
this point. 
The MQT in this ideal condition was studied by two of the authors
(YO and MM).~\cite{2gapOtaMQT}
Figure 3 indicates that the MQT escape rate is enhanced by the JL mode for
various hetero Josephson junctions with two-band superconductors,
whereas the effect of MQT suppression is marginal. 

The results in Fig.~3 indicates that the energy dissipation is relatively
small, compared to the energy of the bounce motion of $\theta$.  
To clarify this point, we calculate the ratio of the dissipative action to
the bounce action. 
Figure 4 shows the contour map of $S_{\rm D}/S_{\rm B}$, with different
junction parameters \mbox{$(\delta \alpha,\,\delta E_{\rm J})$}. 
The open square and the triangle in Fig.~4 indicate the parameter sets
\mbox{$(\delta \alpha,\,\delta E_{\rm J})=(-0.126,\, 0.251)$} and 
$(-0.236,\,0.447)$, respectively. 
The former is evaluated by typical material parameters for
$\textrm{BaFe}_2 \textrm{As}_2$,~\cite{IronPara,ParaCalc} while the latter for
$\textrm{MgB}_2$.~\cite{MgPara}
We find that $S_{\rm D}$ is much smaller than $S_{\rm B}$.
%indeed, $S_{\rm D}/S_{\rm B}$ is smaller than 0.1 for various parameter sets. 
It is noteworthy that the environment $\psi$ oscillates with single 
angular frequency $\omega_{\rm L}$. 
Thus, there is only one dissipation channel in our system. 
This fact would lead to a small energy dissipation. 

Finally, for clarifying the dependence of $\Gamma$ on the inter-band
Josephson energy, we calculate $\Gamma$, with different $E_{\rm in}/E_J$. 
Figure 5(a) shows $\Gamma$ as a function of $E_{\rm in}/E_{\rm J}$.
In this calculation, we use again the typical material parameters for
$\textrm{BaFe}_2 \textrm{As}_2$ and $\textrm{MgB}_2$, as seen in
Fig.~4. 
We find that $\Gamma$ sharply increases with decreasing 
$E_{\rm in}/E_{\rm J}$. 
In order to understand this behavior, 
we plot $\varepsilon$ and $C_{\rm int}$ as functions of 
$E_{\rm in}/E_{\rm J}$ in Figs.~5(b) and (c). 
The magnitude of the zero-point fluctuations of the JL mode
$\varepsilon$ is large, with decreasing $E_{\rm in}/E_{\rm J}$. 
This behavior corresponds to the fact that the JL angular frequency
$\omega_{\rm L}$ decreases when $E_{\rm in}$ decreases. 
Thus, the zero-point fluctuations $\Omega^{2}$ increase, for small
$E_{\rm in}$ [See Eq.~(\ref{mfapsi}), as well].  
Therefore, the reduction of the tunneling barrier height is marked for
small $E_{\rm in}$. 
We also find that the coupling strength of the quantum dissipation 
$C_{\rm int}$ for the $\textrm{MgB}_2$ parameter set is larger than
the $\textrm{BaFe}_2 \textrm{As}_2$ parameter set, as shown in Fig.~5(c). 
In contrast, we find little difference in $\varepsilon$ for these two
parameter sets. 
Hence, the energy dissipation of $\textrm{MgB}_2$ is remarkable,
compared to $\textrm{BaFe}_2 \textrm{As}_2$. 
As a result, the enhancement of the MQT rate for $\textrm{MgB}_2$ is
smaller than that of $\textrm{BaFe}_2 \textrm{As}_2$. 

Let us now summarize our results. 
The MQT in hetero Josephson junctions with two-band superconductors is
strongly affected by the presence of the JP-JL coupling. 
In other words, the MQT escape rate reflects the nature of the JL mode
characterized by the superconducting material parameters, i.e., the
density of states near the interfaces and the inter-band Josephson energy. 

\section{Conclusion}
We constructed a theory of macroscopic quantum tunneling (MQT) in Josephson junctions consisting of
multi-band superconductors, 
and clarified the effect of inter-band phase fluctuations, namely, the Josephson Leggett (JL)
mode on the MQT.
In order to discuss the essential effect of the JL mode,
we employed a minimal model of the multi-band Josephson junction: 
a hetero Josephson junction consisting of a conventional single-gap
superconductor and a two-gap superconductor.
We focused on the in-phase tunneling path along the center-of-mass
motion of the phase differences, which is directly related to 
low-temperature switching events. 
In the tunneling process along the in-phase path, 
the effect of the JL mode is caused by the interaction between the JP
mode and the JL mode.  
We derived a Lagrangian which explicitly includes the JL-JP coupling  
by using a mean-field approximation. 
The derived Lagrangian is similar to that of the Caldeira-Leggett model 
for dissipative quantum tunneling. 
Based on the imaginary-time path-integral method, we derived a formula
for the escape rate from this Lagrangian. 

In our junction, the JL mode plays two major roles which are opposite to
each other: 
(i) the enhancement of quantum tunneling by lowering the tunneling
barrier height, and 
(ii) the suppression of quantum tunneling by quantum dissipation.
We calculated the MQT escape rate, systematically varying the junction parameters. 
We clarified that the enhancement effect is dominant, and that the MQT
escape rate is significantly enhanced by the JL mode. 
The amount of the MQT enhancement depends on the properties of the JL
mode characterized by the superconducting material parameters, such as 
the inter-band Josephson energy and the density of states near the
interfaces of the junctions. 
Therefore, a precise analysis of the MQT would provide valuable
information for the JL mode in multi-band superconductors. 

\section*{Acknowledgements}
 We wish to thank for I. Kakeya and S. Kashiwaya for valuable
 discussions and comments. 
This work is partially supported by a Grant-in-Aid for JSPS Fellows,
a Grant-in-Aid for Scientific Research from the Ministry of Education, Science, Sports and Culture of Japan (Grant No. 24510146).
FN is partially supported by 
RIKEN iTHES Project, MURI Center for Dynamic
Magneto-Optics, JSPS-RFBR Contract No.
12-02-92100,  and a Grant-in-Aid for Scientific Research (S).


\begin{thebibliography}{99}
\bibitem{ParaCalc} 
We can estimate $\delta\! \alpha$ and $\delta\! E_{\rm J}$, with the density
	of states on the Fermi surface, $D_{i}(E_{\rm F})$ for the $i$th
	band. 
We obtain $\alpha_i \propto 1/\sqrt{D_{i}(E_{\rm F})}$ and 
$E_{{\rm J},i} \propto D_{i}(E_{\rm F})$. 
For $\textrm{BaFe}_2 \textrm{As}_2$, we choose the contributions from
	the first and the second largest superconducting gaps of the
	five bands.
\end{thebibliography}
\end{document}